\documentclass[aps,prd,twocolumn,tightenlines,nofootinbib,showpacs]{revtex4-1}
\usepackage[utf8]{inputenc}
\usepackage{amsmath}
\usepackage{amsfonts}
\usepackage{amssymb}
\usepackage{graphicx}
\usepackage[left=2cm,right=2cm]{geometry}
\usepackage{color}
\usepackage[bookmarks=false]{hyperref}
\usepackage[dvipsnames]{xcolor}
\usepackage{ulem}

\newcommand{\eps}{\epsilon}

\begin{document}
\title{Determination of the compositeness of resonances from decays: the case of the $\boldsymbol{B^0_s\to J/\psi f_1(1285)}$}

\author{R. Molina{$^1$}, M. D\"oring$^{1,2}$ and E. Oset$^3$
       }
       \email{oset@ific.uv.es}

\affiliation{
$^1$Institute for Nuclear Studies and Department of Physics, The George Washington University, 725 21st St. NW,
Washington, DC 20052, USA. \\
$^2$ Thomas Jefferson National Accelerator Facility,
12000 Jefferson Ave, Newport News, VA 23606, USA.\\
$^3$Departamento de F\'{\i}sica Te\'orica and IFIC, Centro Mixto Universidad de Valencia-CSIC Institutos de Investigaci\'on de Paterna, Aptdo. 22085, 46071 Valencia, Spain
} 

\begin{abstract}
We develop a method to measure the amount of compositeness of a resonance, mostly made as a bound state of two hadrons, by simultaneously measuring the rate of production of the resonance and the mass distribution of the two hadrons close to threshold. By using different methods of analysis we conclude that the method allows one to extract the value of 1-Z with about $0.1$ of uncertainty. The method is applied to the case of the  $\bar B^0_s \to J/\psi f_1(1285)$ decay, by looking at the resonance production and the mass distribution of $K \bar K^*$.
\end{abstract}

\maketitle

\section{Introduction}
  
One of the recurring questions appearing in the study of hadronic resonances is their internal structure \cite{Klempt:2007cp,Crede:2008vw,Klempt:2009pi}. The simple picture of mesons and baryons being $q \bar q$ and $q q q$, respectively, has given rise to more complicated structures in many cases, with light scalar mesons widely accepted to be some kind of molecular states stemming from the interaction of pseudoscalar mesons  \cite{npa,kaiser,locher,juanarriola}, or the case of the two $\Lambda(1405)$, widely accepted as composite states of $\bar K N$ and $\pi \Sigma$ \cite{cola,Borasoy:2005ie,Oller:2005ig,Oller:2006jw,Borasoy:2006sr,Hyodo:2008xr,Mai:2014xna}, among many others \cite{reviewreso}. 

   One of the pioneer works to determine whether states are composite, or more of the elementary type, is the one of Weinberg, determining that the deuteron is a simple bound state of a proton and a neutron, which gets bound by an interacting potential \cite{compositeness}. The method has been used to determine that some resonances are not elementary, like the $f_0(980)$ and the $a_0(980)$  \cite{kalash}. It relies basically upon determining from experiment the coupling of a state to its assumed components and then making the test of compositeness. In our language $g^2 \partial G/ \partial s=-1$ is the condition to have a composite state \cite{danijuan}. The coupling can be obtained from known scattering amplitudes of the components. In most cases this is difficult and one has only access to certain resonances through decay of heavier ones. This is for instance the decay $B^0_s \to J/\psi f_1(1285)$. The  $f_1(1285)$ is dynamically generated from the interaction of  $K \bar K^*$ as a single channel in the chiral unitary approach \cite{rocasingh}. The inclusion of higher order terms in the Lagrangian barely makes any change in this resonance \cite{Zhou:2014ila} and then reactions producing this resonance become a good testing ground for the method that we propose. Essentially, the method compares two quantities: the production of the resonance, irrelevant of its decay, and the strength of the invariant mass distribution for the production of the assumed molecular components of the resonance. We show that the ratio of these two magnitudes, removing the phase space factors provides a useful information from which the compositeness of the resonances can be determined. 

  In the present paper we develop the formalism and apply it to the case of $\bar{B}^0_s \to J/\psi f_1(1285)$. We show that the method is rather stable with respect to fair changes of some parameters and that one can determine the compositeness with some precision. Experimental information on the rate for this reaction is available from Ref. \cite{Aaij:2013rja}. However, the second part of the information needed in the test, the $K\bar{K}$ invariant mass distribution in $\bar{B}^0_s \to J/\psi K \bar K^*$, is not yet available. The idea exposed in this paper should provide an incentive for measuring such reaction.  

\section{Formalism}
\begin{figure}[h]
\begin{center}
 \includegraphics[width=0.9\linewidth]{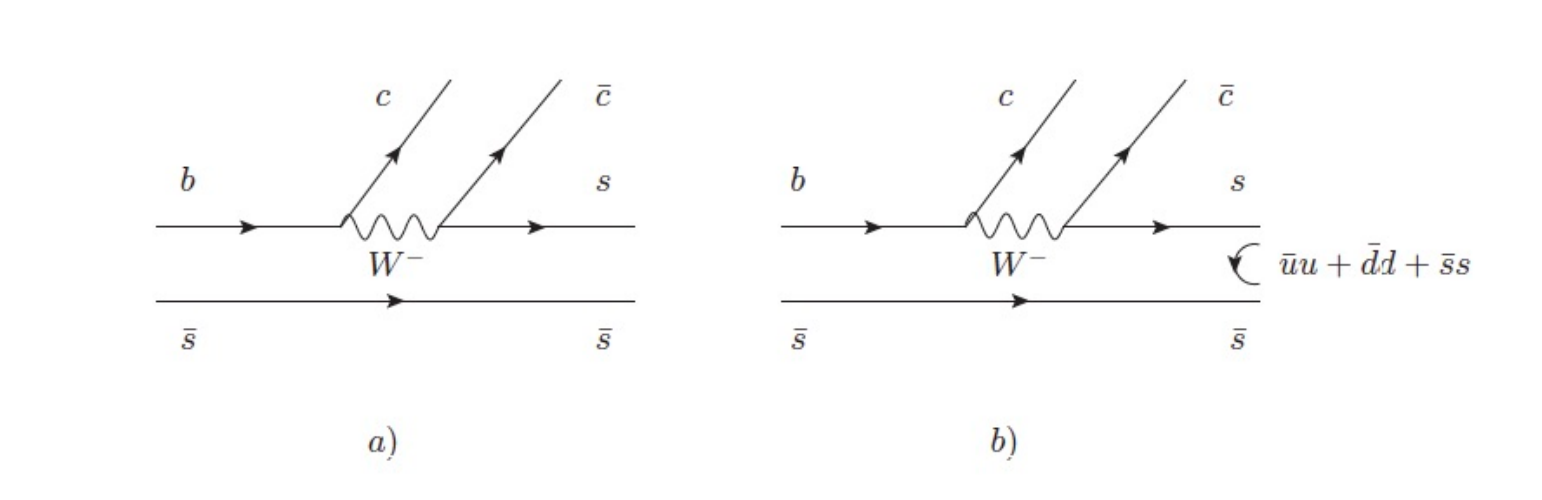}
 \end{center}
 \caption{a) Elementary quark arrangement for the decay. b) Hadronization of  the final $s\bar{s}$ component.}
 \label{fig:fig1}
\end{figure}

In Fig. \ref{fig:fig1} the $\bar{B}^0_s\to J/\psi f_1(1285)$ decay at microscopical level is depicted \cite{Liang:2014tia}.

In order to hadronize the final $s\bar{s}$ component we must bear in mind the molecular structure of the $f_1(1285)$ in our picture. The state is a $0^+(1^{++})$. With the prescription followed here for $C$-parity, such that $CV=-\bar{V}$, with $V$ a vector meson, the representation for the $f_1(1285)$ is 
\begin{equation}
 \vert f_1(1285) \rangle =\frac{1}{2}(K^{*+}K^-+K^{*0}\bar{K}^0-K^{*-}K^+-\bar{K}^{*0}K^0)\ .\label{eq:f1}
\end{equation}
Let us see how this state emerges from the mechanism of Fig. \ref{fig:fig1}b). In the hadronization we will include a $\bar{q}q$ pair with the quantum numbers of the vacuum, $\bar{u}u+\bar{d}d+\bar{s}s$, and will obtain two pairs of $q\bar{q}$. One of them will correspond to a vector meson and the other one to a pseudoscalar. In terms of pseudoscalars and vectors the $q\bar{q}$ matrix $M$,
\begin{eqnarray}
M=\begin{pmatrix}u\bar{u}& u\bar{d}& u\bar{s}\\d\bar{u}&d\bar{d}&d\bar{s}\\s\bar{u}&s\bar{d}&s\bar{s}\end{pmatrix}\ ,
\end{eqnarray}
can be written as
\begin{eqnarray}
 &&M_P=\nonumber\\&&\begin{pmatrix}
      \frac{\pi^0}{\sqrt{2}}+\frac{\eta}{\sqrt{3}}+\frac{\eta'}{\sqrt{6}}&\pi^+&K^+\\
      \pi^-&-\frac{\pi^0}{\sqrt{2}}+\frac{\eta}{\sqrt{3}}+\frac{\eta'}{\sqrt{6}}&K^0\\
      K^-&\bar{K}^0&-\frac{1}{\sqrt{3}}\eta+\sqrt{\frac{2}{3}}\eta'
     \end{pmatrix}\ ,\nonumber\\
\end{eqnarray}
which implements the standard $\eta-\eta'$ mixing \cite{Bramon:1992kr}, and
\begin{eqnarray}
 M_V=\begin{pmatrix}
      \frac{\rho^0}{\sqrt{2}}+\frac{\omega}{\sqrt{2}}&\rho^+&K^{*+}\\
      \rho^-&-\frac{\rho^0}{\sqrt{2}}+\frac{\omega}{\sqrt{2}}&K^{*0}\\
      K^{*-}&\bar{K}^{*0}&\phi
     \end{pmatrix}\ .\nonumber\\
\end{eqnarray}
The matrix $M$ obeys
\begin{eqnarray}
 MM=M(\bar{u}u+\bar{d}d+\bar{s}s)\ .
\end{eqnarray}
Hence, a hadronized $q\bar{q}$ component can be written in terms of matrix elements of $MM$ and then as matrix
elements of $M_VM_P$ or $M_PM_V$. It is easy to see that the combination 
\begin{equation}
 (M_VM_P\pm M_P M_V)
\end{equation}
has $C$ parity $-$ and $+$ respectively. The $C=+$ state (up to normalization), will be
\begin{eqnarray}
 &&s\bar{s}(\bar{u}u+\bar{d}d+\bar{s}s)=M_{33}(\bar{u}u+\bar{d}d+\bar{s}s)\equiv (MM)_{33}\nonumber\\&&\longrightarrow
 (M_P M_V-M_V M_P)_{33}=\nonumber\\&&K^-\bar{K}^{*+}+\bar{K}^0K^{*0}+(-\sqrt{\frac{1}{3}}\eta+\sqrt{\frac{2}{3}}\eta')\phi\nonumber\\&&
-K^{*-}K^+-\bar{K}^{*0}K^0-\phi(-\frac{1}{\sqrt{3}}\eta+\sqrt{\frac{2}{3}}\eta')\ .\label{eq:qq}
 \end{eqnarray}
We can see that the $\phi$ term cancels because $C(\phi\eta)=(-)(+)=(-)$. The combination in Eq. (\ref{eq:qq}) is then the same one as in Eq. (\ref{eq:f1}) and it is also projected
over $I=0$ because it comes from $s\bar{s}$ after hadronization with $q\bar{q}$ with the quantum
numbers of the vacuum.

\section{Coalescence production of the $\boldsymbol{f_1(1285)}$}
\begin{figure}[h]
 \begin{center}
  \includegraphics[width=0.7\linewidth]{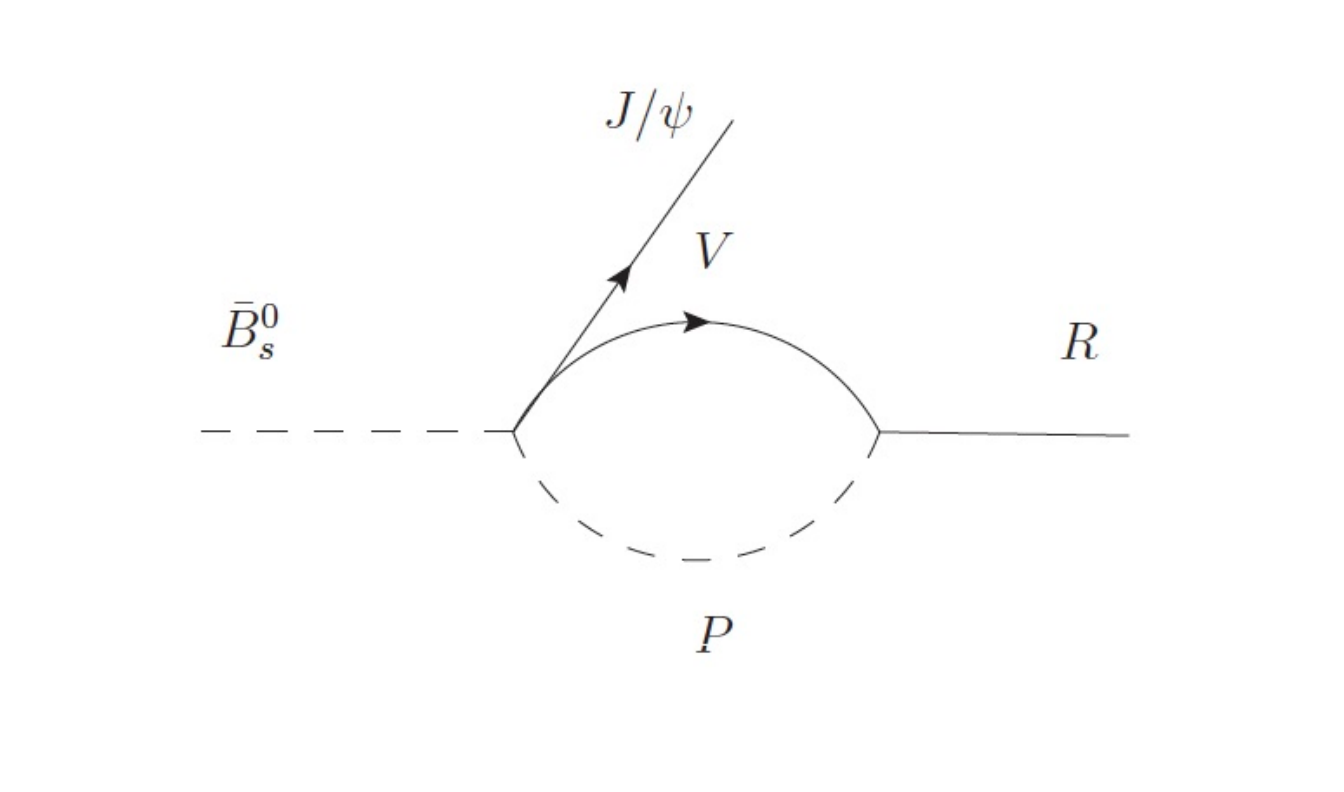}
 \end{center}
\caption{Diagrammatic representation of the production of a molecular resonance $R$ in the process $\bar{B}^0_s\to J/\psi R$.}
\label{fig:fig2}
\end{figure}

In this section we write the formalism for the $\bar{B}^0_s\to J/\psi f_1(1285)$ decay, irrelevant on how the $f_1(1285)$ decays later. Diagrammatically it is represented in Fig. \ref{fig:fig2},

The idea is that being the $f_1(1285)$ a molecule, to form it we must produce the $VP$ components which merge into the resonance. This process, evaluated at the bound state energy $s=s_R$ is given by
\begin{equation}
 t(\bar{B}^0_s\to J/\psi f_1(1285))=V_P G_{K\bar{K}^*}(s_R)g_{R,K\bar{K}^*}\label{eq:t}
\end{equation}
where $G_{K\bar{K}^*}$ is the loop function of the intermediate propagator of $K$ and $\bar{K}^*$ and $g_{R,K\bar{K}^*}$ stands for the coupling of the 
$f_1(1285)$ to the $K\bar{K}^*-\bar{K}K^*$ component of Eq. (\ref{eq:f1}) with that normalization. Recall that $g_{R, K\bar K^*}^2$ is the residue at the pole. In the following we drop the index of $g$. $V_P$ in Eq. (\ref{eq:t}) factorizes the weak and the hadronization processes in the 
relatively narrow region of energies from the mass of the $f_1(1285)$ to about $200$ MeV above the $K\bar{K}^*$ threshold. $V_P$ is unknown, and in some works it is given in terms of form factors 
\cite{hanhart,weiwang}, which ultimately are parametrized to some data. We assume $V_P$ constant in the range of energies that we study, which finds support in the works of \cite{hanhart,
Kang:2013jaa}. Our strategy is to cancel the factor in some ratios for which we can make predictions with no free parameters. Note also that direct resonance production without a $VP$ intermediate state is possible. The relevance of this contribution is
discussed in Sec. IX.
\section{$\boldsymbol{\bar{B}^0_s\to J/\psi(K\bar{K}^*,\bar{K}K^*)}$ decay}
\begin{figure}
 \begin{center}
  \includegraphics[width=1\linewidth]{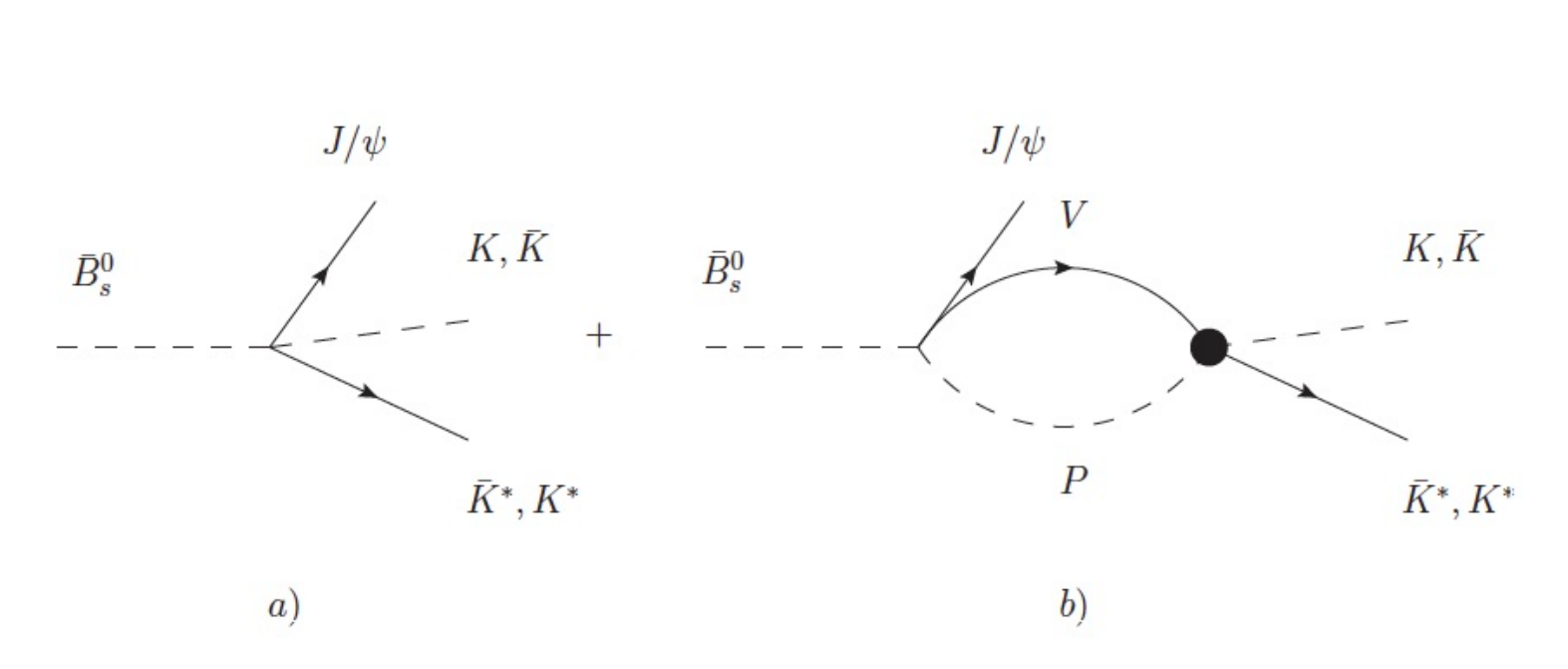}
 \end{center}
\caption{Diagrammatic representation for $K\bar{K}^*$, $\bar{K}K^*$ production. a) Tree level, b) rescattering. Here the $V$ and $P$ represent vector and pseudoscalar meson, respectively.}
\label{fig:fig3}
\end{figure}

 The $f_1(1285)$ is bound
by about $100$ MeV with respect to the $K\bar{K}^*$ threshold, hence by looking at the $K\bar{K}^*$ production we shall not see the peak, but just the tail of the resonance. Diagrammatically, the process proceeds as in Fig. \ref{fig:fig3}.
Note that for resonance creation shown in Fig. \ref{fig:fig2}, the intermediate $K\bar{K}^*,\bar{K}K^*$ states merges into the $f_1(1285)$ resonance. In contrast, the $K\bar{K}^*,\bar{K}K^*$ final state of Eq. (\ref{eq:f1}) can also be produced at tree level (Fig. \ref{fig:fig3} a)), apart from the rescattering mechanism (Fig. \ref{fig:fig3} b)). Analytically, we have
\begin{equation}
 t(\bar{B}^0_s\to J/\psi K\bar{K}^*)=V_P(1+G_{K\bar{K}^*}(s)t_{K\bar{K}^*,K\bar{K}^*}(s))
\end{equation}
where $K\bar{K}^*$ refers to the combination of Eq. (\ref{eq:f1}) and $t_{K\bar{K}^*,K\bar{K^*}}$ stands for the scattering matrix of the normalized state of Eq. (\ref{eq:f1}). In practice, assuming one knows that the positive $C$-parity state is produced, an experimentalist will measure any of the four components of Eq. (\ref{eq:f1}), each of
which has a probability $\frac{1}{4}$ with respect to the production of that normalized combination. The sum of the four components would give the same result as the production of the normalized component of Eq. (\ref{eq:f1}) that we are calculating. The idea is to construct a ratio of the two rates of the mechanisms of Figs. \ref{fig:fig2} and \ref{fig:fig3} in order to cancel the unknown factor $V_P$ and make predictions which are tied to the nature of this resonance.

The decay rate for $\bar{B}^0_s\to J/\psi f_1(1285)$, which is known experimentally \cite{Aaij:2013rja}, is given in terms of $t$ of Eq. (\ref{eq:t}) as
\begin{equation}
 \Gamma(\bar{B}^0_s\to J/\psi f_1(1285))=\frac{1}{8 \pi}\frac{1}{m^2_{\bar{B}^0_s}}p_{J/\psi,p}\vert t(\bar{B}^0_s\to J/\psi f_1)\vert^2
\end{equation}
where $p_{J/\psi,p}$ is the momentum of the $J/\psi$ in the rest frame of the $\bar{B}^0_s$, calculated at the pole of the $f_1(1285)$.

On the other hand, we obtain the invariant mass distribution for the $\bar{B}^0_s\to J/\psi K\bar{K}^*$ decay by means of
\begin{eqnarray}
&&\frac{d \Gamma(\bar{B}^0_s\to J/\psi K\bar{K}^*)}{d M_{\mathrm{inv}}}=\nonumber\\&&\frac{1}{(2\pi)^3}\frac{p_{J/\psi}\tilde{p}_K}{4 m^2_{\bar{B}^0_s}}\vert t(\bar{B}^0_s\to J/\psi K\bar{K}^*)(M_\mathrm{inv})\vert^2\nonumber\\\label{eq:rat}
\end{eqnarray}
where $p_{J/\psi}$ is now the $J/\psi$ momentum for the $\bar{B}^0_s\to J/\psi K\bar{K}^*(M_\mathrm{inv})$. The $K\bar{K}^*$ or $\bar{K}K^*$ states in Eq. (\ref{eq:rat}) have an invariant mass $M_\mathrm{inv}$, and $\tilde{p}_K$ is the $K$ momentum in the $K\bar{K}^*$ rest frame. Following the suggestion of Ref. \cite{Liang:2015twa}, we define a dimensionless quantity, where we have also removed the phase space factors in $\frac{d\Gamma}{dM_{\mathrm{inv}}}$,
\begin{eqnarray}
&& \frac{dR_\Gamma}{dM_{\mathrm{inv}}}=\nonumber\\&&\frac{1}{\Gamma(\bar{B}^0_s\to J/\psi f_1(1285))}\frac{s^{3/2}_R}{p_{J/\psi}\tilde{p}_K}\frac{d\Gamma(\bar{B}^0_s\to J/\psi K\bar{K}^*)}{dM_\mathrm{inv}}\nonumber\\&&
=\frac{1}{4\pi^2}\frac{s^{3/2}_R}{p_{J/\psi,p}}\left| \frac{t(\bar{B}^0_s\to J/\psi K\bar{K}^*)}{t(\bar{B}^0_s\to J/\psi f_1(1285))}\right|^2\nonumber\\&&
=\frac{1}{4\pi^2}\frac{s^{3/2}_R}{p_{J/\psi,p}}\left|\frac{1+G_{K\bar{K}^*}(M_{\mathrm{inv}}^2)t_{K\bar{K}^*,K\bar{K}^*}}{gG_{K\bar{K}^*}(s_R)}\right|^2\ .\label{eq:rat2}
\end{eqnarray}
\section{The chiral unitary model for the $\boldsymbol{f_1(1285)}$}\label{sec:chroca}
\begin{figure}
 \begin{center}
  \includegraphics[width=0.8\linewidth]{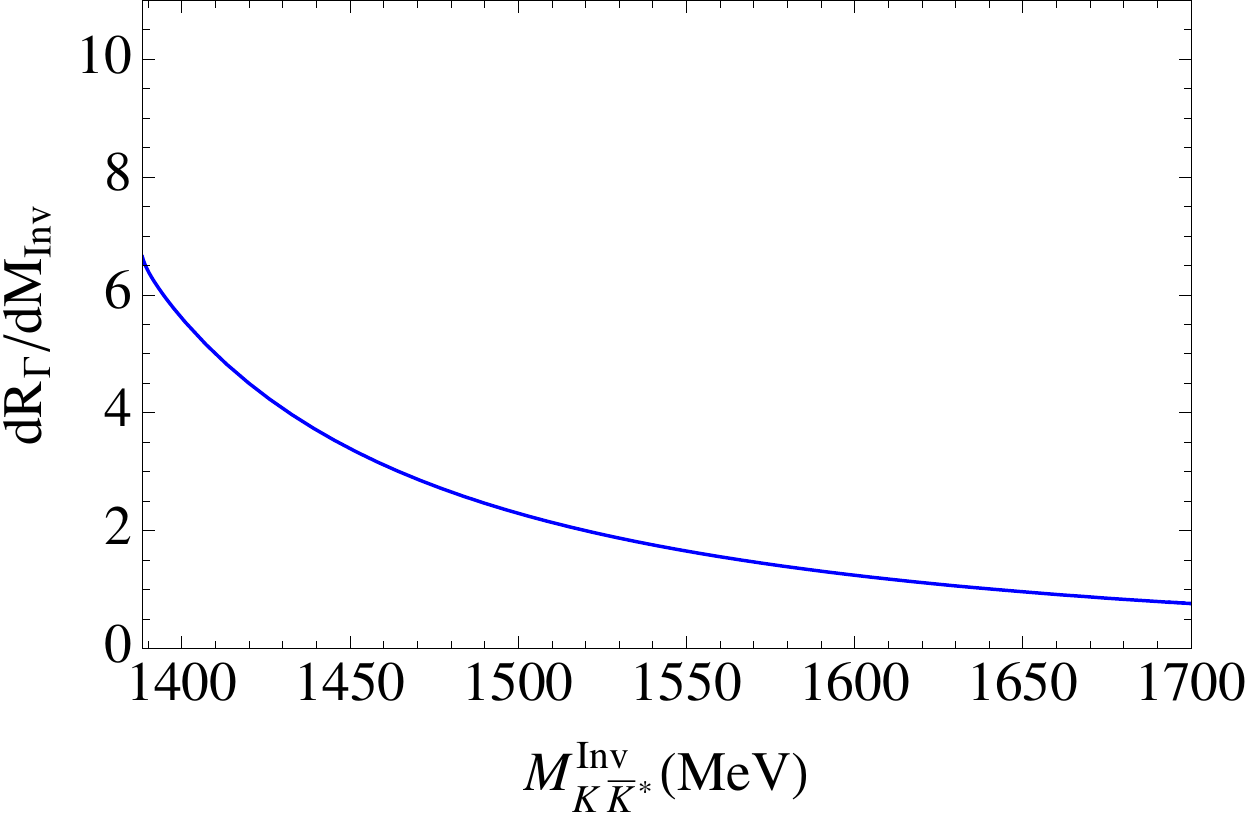}
 \end{center}
\caption{$\frac{dR_\Gamma}{dM_\mathrm{inv}}$ from Eq. (\ref{eq:rat2}) as a function of $M_\mathrm{inv}$.}
\label{fig:fig4}
\end{figure}

To illustrate the method to determine compositeness, an explicit model for the $f_1(1285)$ is studied. We follow here the chiral unitary approach of Ref. \cite{rocasingh}, based on the chiral Lagrangian \cite{birse}. This
chiral Lagrangian is readily obtained using the local hidden gauge approach \cite{hidden1,hidden2,hidden4}, exchanging vector
mesons between the $K$ and the $K^*$ and neglecting the square of the transfered four-momentum versus the square of the vector-meson mass. Higher-order terms are considered in Ref. \cite{Zhou:2014ila} but the changes in this channel are minimal compared to the lowest order.
The $S$-wave potential obtained in Ref. \cite{rocasingh} is of the type $V\vec{\eps}\cdot \vec{\eps}\,'$, where $\vec{\eps}\cdot\vec{\eps}\,'$ are the polarization vectors of the initial and final vectors and $V$ is given by 
\begin{eqnarray}
 && V=-\frac{1}{8f^2}3\left[ 3 s-(M^2+m^2+M'^2+m'^2)\right.\nonumber\\&&\left.-\frac{1}{s}(M^2-m^2)(M'^2-m'^2)\right]\label{eq:rocap}
\end{eqnarray}
where $M,M'=M_{K^*}$, $m=m'=m_K$ and $f=f_\pi=93\,\mathrm{MeV}$, and $s$ should be identified with $M^2_{\mathrm{inv}}$ of Eq. (\ref{eq:rat}). The scattering matrix, of the type $T\vec{\eps}\cdot\vec{\eps}\,'$, is given by
\begin{eqnarray}
 T=\left[1-VG\right]^{-1}V=\left[V^{-1}-G\right]^{-1}\ .\label{eq:bethe}
\end{eqnarray}
The potential of Eq. (\ref{eq:rocap}) is attractive and leads to a bound state. We then use a cutoff method to regularize the $G$ function which becomes
\begin{eqnarray}
 &&G(s)=\nonumber\\&&\int_{0}^{q_\mathrm{max}}\frac{q^2dq}{2\pi^2}\frac{\omega_1(q)+\omega_2(q)}{2\omega_1(q)\omega_2(q)}\frac{1}{s-(\omega_1(q)+\omega_2(q))^2+i\eps}\ ,\nonumber\\
\end{eqnarray}
where $\omega_{1,2}(q)=\sqrt{\vec{q}\,^2+m^2_{1,2}}$, and $1,2$ stands for $K,K^*$. We then fix $q_\mathrm{max}$ in order to get the bound state at $\sqrt{s}_R=1285$ MeV, obtaining $q_\mathrm{max}=950$ MeV, which is a value of natural size.
We next calculate $g$ given by
\begin{eqnarray}
 g^2=\mathrm{lim}_{s\to s_R}(s-s_R)T=\frac{1}{\frac{\partial{V}^{-1}}{\partial s}\vert_{s_R}-\frac{\partial G}{\partial s}\vert_{s_R}}\label{eq:g2}
\end{eqnarray}
where L'H\^{o}pital's rule has been applied in the last step. We thus have all the elements to evaluate $\frac{d R_\Gamma}{dM_\mathrm{inv}}$ of Eq. (\ref{eq:rat2}) and show the results in Fig. \ref{fig:fig4}.

Note that $\frac{dR_\Gamma}{dM_\mathrm{inv}}$ decreases with increasing invariant mass near threshold. Should we have evaluated $\frac{d\Gamma}{dM_\mathrm{inv}}$ it would have an accumulation of strength around threshold, due to the existence of the resonance below threshold. However, in $\frac{dR_\Gamma}{dM_\mathrm{inv}}$ we have divided by the phase space factors
$p_{J/\psi}\tilde{p}_K$, and then, up to some factors, what we see in Fig. \ref{fig:fig4} is essentially
\begin{eqnarray}
 \frac{d R_\Gamma}{d M_\mathrm{inv}}\sim \left(\frac{1}{s-s_R}g\right)^2\ .\label{eq:coup}
\end{eqnarray}
Hence, this quantity has two important elements. Its shape reveals that there is a resonance below threshold. Its strength is related to the coupling 
of the resonance to the $K\bar{K}^*-\bar{K}^*K$ state. The coupling contains information concerning the nature of the $f_1(1285)$ resonance, through the method which we briefly discuss below.
\section{Compositeness and elementariness of resonances}
In a reknown paper \cite{compositeness} Weinberg studied $np$ scattering and determined that the deuteron was not an ``elementary'' particle but a ``composite'' one made from a neutron and a proton interacting through
the potential responsible for the scattering properties of $np$ at small energies. One could envisage an opposite extreme in which the deuteron could be a compact object of six quarks with practically no coupling to $np$, which we would accept as ``elementary''. The coupling of the deuteron to the $np$ component is what describes the amount of compositeness, and the 
method of Weinberg determines that quantitatively for lightly bound systems in $s$-wave. Application of the method to different cases is done in \cite{kalash}. The method is generalized to the case of more bound states and coupled channels in \cite{danijuan}, to resonances in \cite{yamagata} and to other partial waves in \cite{aceti}. Different derivations and reformulations are given in a series of papers \cite{jido, hyodo,sekihara}. In the present case with one channel and one bound state we can reformulate the issue in a very simple way. Let us start from Eq. (\ref{eq:g2}). We find 
\begin{eqnarray}
 -g^2\frac{\partial G}{\partial s}+g^2\frac{\partial V^{-1}}{\partial s}\equiv 1\ .\label{eq:com}
\end{eqnarray}
According to Refs. \cite{jido, hyodo,sekihara} the first term in Eq. (\ref{eq:com}), $1-Z=-g^2\frac{\partial G}{\partial s}$, measures the 
compositeness of the state (in this case as made of ($K\bar{K}^*-\bar{K}K^*)$), while the second term $Z=g^2\frac{\partial V^{-1}}{\partial s}$ is the amount of non-composite nature, which is often called elementariness \cite{jido, hyodo,sekihara,juancarmen}, but, as discussed in 
Ref. \cite{acetidai}, it also accounts for other composite channels not explicitly included in the basis that one has chosen to describe the state. Indeed, in Ref. \cite{acetidai} it is explicitly shown how, starting from two channels and energy independent potentials, one can eliminate one channel and still describe the other one by means of an effective potential, which, however, becomes energy dependent. The second term in Eq. (\ref{eq:com}), $g^2\frac{\partial V^{-1}}{\partial s}$, gives in this case the probability of
the channel that we have eliminated. In any case, our aim here is to extract, from an experimental measurement of $\frac{d R_\Gamma}{d M_{\mathrm{inv}}}$, the probability to have $K\bar{K}^*-\bar{K}^*K$ in the wave function of the $f_1(1285)$, which will be given by $-g^2\frac{\partial G}{\partial s}$. Since in $\frac{dR_\Gamma}{dM_\mathrm{inv}}$ there is information about $g^2$ (see Eq. (\ref{eq:coup})), it is clear that one can get information on $1-Z$, or $Z$ from $\frac{d R_\Gamma}{d M_\mathrm{inv}}$. How to do that without the need to know the explicit form of the potential $V$ and the value of the cutoff to regularize $G$, is shown in the next section. We will use two forms of potentials to estimate systematic uncertainties.
\section{Analysis of $\boldsymbol{d R_\Gamma /d M_\mathrm{inv}}$}
\subsection{Linear potential}
The potential of Eq. (\ref{eq:rocap}), by an expansion around the pole, becomes
\begin{eqnarray}
 V=V(s_R)+\beta \frac{s-s_R}{M^2_R}\label{eq:lpot}
\end{eqnarray}
up to linear terms in $s$. The condition that Eq. (\ref{eq:bethe}) has a pole at $s_R$ can be written as
\begin{equation}
 V^{-1}(s_R)-G(s_R)=0\ .\label{eq:20}
\end{equation}
Then,
\begin{equation}
 \left.\frac{\partial V^{-1}}{\partial s}\right|_{s_R}=-\left.\frac{1}{V^2}\frac{\partial V}{\partial s}\right|_{s_R}\label{eq:21}
\end{equation}
and from Eq. (\ref{eq:lpot}) we have
\begin{eqnarray}
 \left.\frac{\partial V^{-1}}{\partial s}\right|_{s_R}=-\left.\frac{1}{V^2}\right|_{s_R}\frac{\beta}{M^2_R}=-G^2(s_R)\frac{\beta}{M^2_R}\ .
\end{eqnarray}
Eq. (\ref{eq:com}) is now rewritten as
\begin{eqnarray}
 -g^2\frac{\partial G}{\partial s}-g^2G(s_R)^2\frac{\beta}{M^2_R}=1\ ,\label{eq:com2}
\end{eqnarray}
where the first term is $1-Z$ and the second one $Z$, which implies $\beta$ negative for physical solutions. 

Our aim now is to write $\frac{\partial R_\Gamma}{\partial M_{\mathrm{inv}}}$ in terms of $Z$ and quantities which can be calculated without knowing the potential and
the regulator of $G$, such that from the measurement of $\frac{ d R_\Gamma}{d M_{\mathrm{inv}}}$ one can determine $Z$ without the need of a model to interpret it. For this 
we write $\frac{|1+Gt|^2}{|gG|^2}$ in terms of $Z$ and calculable quantities. The first step is to eliminate $\beta$ of Eq. (\ref{eq:lpot}) in terms of $Z$. For that purpose, recall from Eq. (\ref{eq:com}) that
\begin{eqnarray}
 \frac{-g^2G^2(s_R)\beta/M^2_R}{\left.-g^2\frac{\partial G}{\partial s}\right|_{s_R}}=\frac{Z}{1-Z}\ ,
\end{eqnarray}
from where 
\begin{equation}
\beta/M^2_R=\left.G(s_R)^{-2}\frac{\partial G}{\partial s}\right|_{s_R}\frac{Z}{1-Z} \ .
\end{equation}
The latter equation allows to obtain $\beta$ in terms of $G$ and $\partial G/\partial s$, but this is not of much help since $G$ needs an unknown regulator. Yet, the factor that we want for $d R_\Gamma/d M_{\mathrm{inv}}$ in Eq. (\ref{eq:rat2}) can be rewritten taking into account that 
\begin{eqnarray}
 1+GT=\frac{T}{V}=\frac{1}{V}\frac{1}{V^{-1}-G}=\frac{1}{1-VG}\ ,
\end{eqnarray}
and we can write
\begin{eqnarray}
 &&\left|\frac{1+GT}{gG(s_R)}\right|^2=\frac{1}{g^2G^2(s_R)}\left|\frac{1}{1-(G(s_R)^{-1}+\beta\frac{s-s_R}{M^2_R})G}\right|^2\nonumber\\&&=
 \left.-\frac{\partial G}{\partial s}\right|_{s_R}\frac{1}{1-Z}\times\nonumber\\&&\left|\frac{1}{G(s_R)-G(s)-G(s_R)^{-1}G(s)\left.\frac{\partial G}{\partial s}\right|_{s_R}\frac{Z}{1-Z}(s-s_R)}\right|^2\ .\nonumber\label{eq:rat3}\\
\end{eqnarray}
This equation is most appropiate because $G$ is logarithmically divergent and must be regularized, but $\partial G/\partial s$ is convergent. We can then use values of $q_\mathrm{max}$ around $1$ GeV and see the 
stability of the results to make a claim of weak model dependence. We also have the term $G(s_R)-G(s)$, which is again convergent, and $G(s_R)^{-1}G(s)$ goes to unity as $q_\mathrm{max}\to \infty$, which means that $G(s_R)^{-1}G(s)$ is smoothly dependent on the cutoff and multiplies a term in Eq. (\ref{eq:rat3}) 
which should be small compared to $G(s_R)-G(s)$, certainly for small values of $Z$. In any case, we shall test the stability of the results by changing $q_\mathrm{max}$ in a reasonable range of natural values.
\subsection{Analysis with a CDD pole}
In order to take into account possible ``elementary'' components, often an analysis using a CDD pole \cite{castillejo} is performed \cite{nsd,sasa}. Assume that the potential is of the type
\begin{equation}
 V=V_0+\gamma \frac{1}{s-s_{\mathrm{CDD}}}\ ,\label{eq:vcdd}
\end{equation}
where $s_{\mathrm{CDD}}$ accounts for a ``bare'' pole of a possible elementary component. The condition of a pole at $s=s_R$ implies
\begin{eqnarray}
 &&V=\nonumber\\&&V_0+\frac{\gamma}{s-s_{\mathrm{CDD}}}+V_0+\frac{\gamma}{s_R-s_{\mathrm{CDD}}}-V_0-\frac{\gamma}{s_R-s_{\mathrm{CDD}}}\nonumber\\
 &&=G(s_R)^{-1}+\gamma\left(\frac{1}{s-s_{\mathrm{CDD}}}-\frac{1}{s_R-s_{\mathrm{CDD}}}\right)\nonumber\\&&=G(s_R)^{-1}+\gamma \frac{s_R-s}{(s-s_{\mathrm{CDD}})(s_R-s_{\mathrm{CDD}})}\ ,
\end{eqnarray}
where we have used that $V_0+\frac{\gamma}{(s_R-s_{\mathrm{CDD}})}=G(s_R)^{-1}$ at the pole. 

We can write from Eqs. (\ref{eq:20}), (\ref{eq:21}) and (\ref{eq:vcdd}),
\begin{eqnarray}
 \left.\frac{\partial V^{-1}}{\partial s}\right|_{s_R}=G(s_R)^2\frac{\gamma}{(s_R-s_{\mathrm{CDD}})^2}\ ,
\end{eqnarray}
and the sum rule of Eq. (\ref{eq:com2}) becomes
\begin{eqnarray}
 \left.-g^2\frac{\partial G}{\partial s}\right|_{s_R}+g^2G(s_R)^2\frac{\gamma}{(s_R-s_{\mathrm{CDD}})^2}=1\ ,\label{eq:com3}
\end{eqnarray}
where the first term stands for $1-Z$ and the second one for Z, where now $\gamma$ will be positive for physical solutions.
Eliminating $\gamma$ in terms of $Z$ and proceeding like in the former subsection, we obtain 
{\small{\begin{eqnarray}
 &&\left|\frac{1+GT}{gG(s_R)}\right|^2=\left.\frac{-\partial G}{\partial s}\right|_{s_R}\frac{1}{1-Z}\times\nonumber\\
 &&\left|\frac{1}{G(s_R)-G(s)-G(s_R)^{-1}G(s)\left.\frac{\partial G}{\partial s}\right|_{s_R}\frac{(s-s_R)(s_R-s_{\mathrm{CDD}})}{(s-s_{\mathrm{CDD}})}\frac{Z}{1-Z}}\right|^2\ .\nonumber\\
\label{eq:rat4}
 \end{eqnarray}}}
Compared to Eq. (\ref{eq:rat3}), Eq. (\ref{eq:rat4}) has the extra factor $(s_R-s_{\mathrm{CDD}})/(s-s_{\mathrm{CDD}})$ in the last term of the denominator. As far as this term is relatively smaller than $G(s_R)-G(s)$ and $s_{\mathrm{CDD}}$ is relatively far from threshold, this extra term will have no much relevance and we can get about the same value of $Z$ as with the other method from an analysis of the experimental data. The value of $s_{\mathrm{CDD}}$ is in principle unknown in the analysis. One can estimate systematic uncertainties from varying its value.
Sometimes, one has information about the range of possible values as we discuss in the next section. And finally it should be noted
that when $s_{\mathrm{CDD}}\to\infty$ we recover the results of Eq. (\ref{eq:rat3}). Certainly, if $Z$ is close to zero, this second term is also very small
and we expect the same result from Eq. (\ref{eq:rat3}) and (\ref{eq:rat4}). In the next section we study the sensitivity of the results to the used method and to the cutoff $q_\mathrm{max}$.

\section{Results}
\begin{figure}[h]
 \begin{center}
  \includegraphics[width=0.8\linewidth]{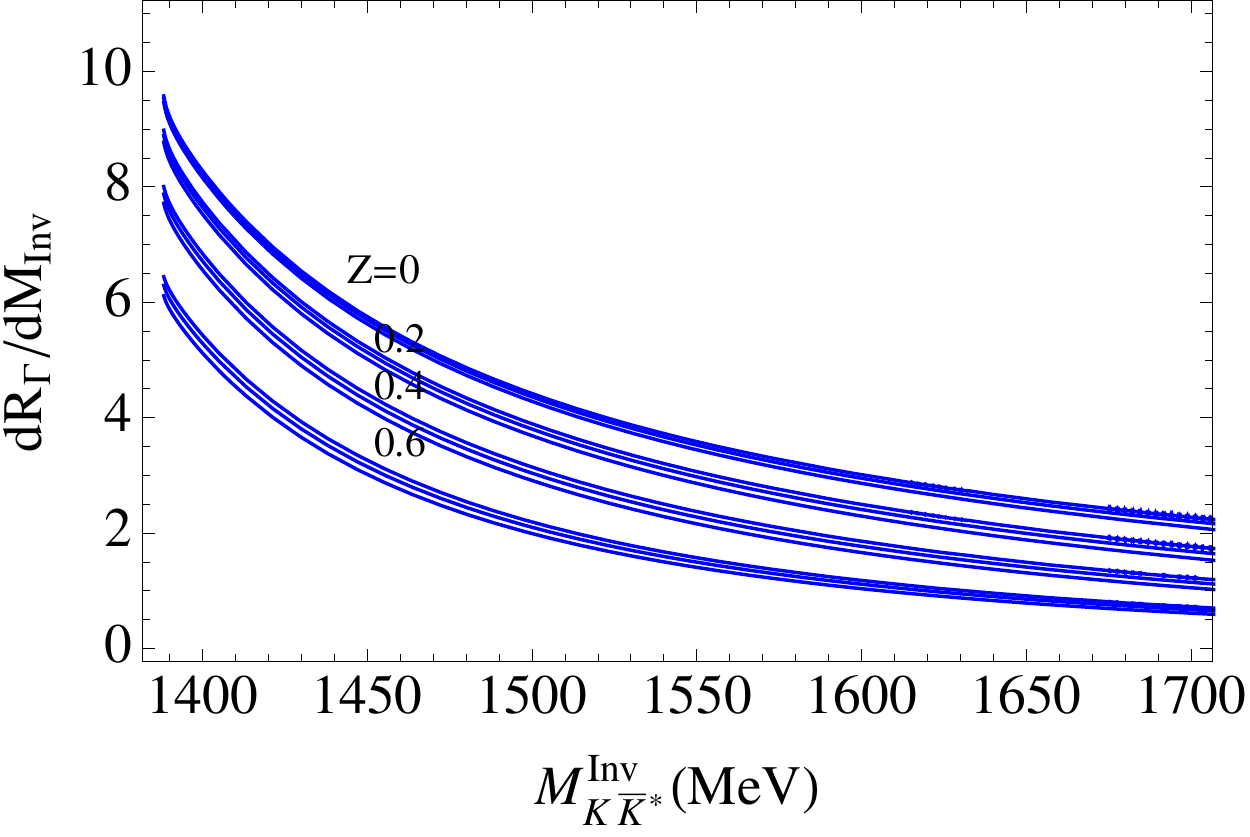}
 \end{center}
\caption{$d R_\Gamma/d M_\mathrm{inv}$ with the potential from Eq. (\ref{eq:lpot}) as a function of $M_\mathrm{inv}$ for several $Z$ values and different values of $q_\mathrm{max}=850,950$ and $1050$ MeV.}
\label{fig:fig5}
\end{figure}

In all figures the $f_1(1285)$ is fixed at $s_R=(1285\,\mathrm{ MeV})^2$. In Fig. \ref{fig:fig5} we show $d R_\Gamma/d M_\mathrm{inv}$ for different values of $Z$ as a function of $M_\mathrm{inv}$, using the linear potential of Eq. (\ref{eq:lpot}) and Eq. (\ref{eq:rat3}). We test $q_\mathrm{max}=850,950$ and $1050$ MeV, a wide range of values of natural size. Recall that the value of $q_\mathrm{max}$ used in the chiral unitary approach of Section \ref{sec:chroca} was $q_\mathrm{max}=950$ MeV, which is within the chosen range. We observe that the results
for $dR_\Gamma/d M_\mathrm{inv}$ barely depend on the value of $q_\mathrm{max}$, as we anticipated, in view of the fact that $ dR_\Gamma/dM_\mathrm{inv}$ in Eq. (\ref{eq:rat3}) only depends on $\partial G/\partial s$ and $G(s_R)-G(s)$, which are both convergent.
\begin{figure}[h]
 \begin{center}
  \includegraphics[width=0.8\linewidth]{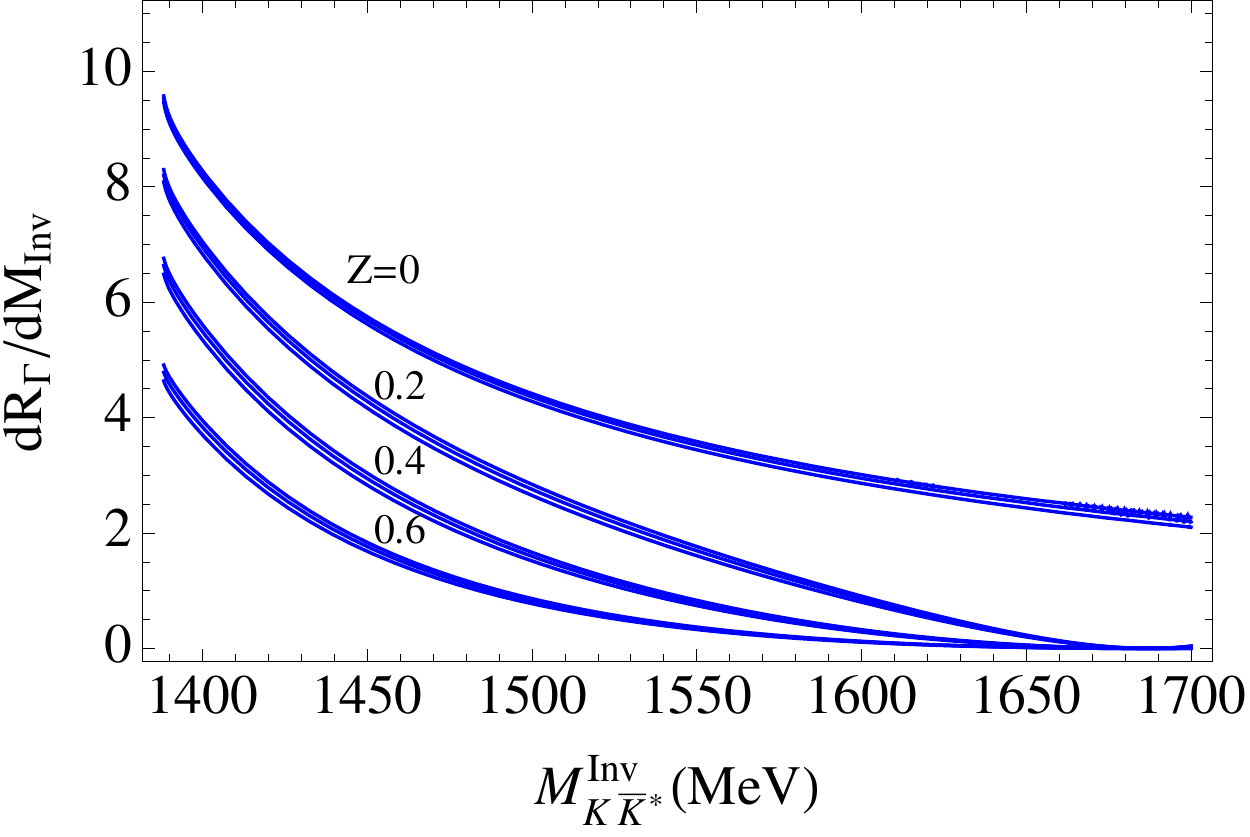}
 \end{center}
\caption{$d R_\Gamma/d M_\mathrm{inv}$ with the potential from Eq. (\ref{eq:vcdd}) as a function of $M_\mathrm{inv}$ for several $Z$ values and different values of $q_\mathrm{max}=850,950$ and $1050$ MeV.}
\label{fig:fig7}
\end{figure}
We can see that the dispersion of the results when changing the cutoff is a bit bigger for $Z\neq 0$, but even then the band of values is narrow enough such that we can differentiate between $Z=0$ and $Z=0.2$, and certainly between $Z=0$ and $Z=0.6$. Note that for $Z=0$ the value of $d R_\Gamma/d M_\mathrm{inv}$ at threshold is about $9.5$ while for $Z=0.6$ it is about $6$. Such difference is clearly visible in an experiment with present statistics. The ratio $dR_\Gamma/dM_{\mathrm{inv}}$ is bigger for larger compositeness $1-Z$. This is expected from its definition given in Eq. (\ref{eq:rat2}), i.e., as the ratio between
the resonance decay into its dynamical $K\bar K^*$ components over its decay into any final state.

Fig. \ref{fig:fig7} shows $dR_\Gamma/dM_\mathrm{inv}$ evaluated with the CDD potential of Eq. (\ref{eq:vcdd}). We see again that the dispersion of the results by varying $q_\mathrm{max}$ is small and similar to the previously discussed method. For states which are mostly molecular,
$\sqrt{s_{\mathrm{CDD}}}$ is far away from threshold. We have chosen $\sqrt{s_{\mathrm{CDD}}}=M_K+M_{K^*}+300$ MeV. The choice is based on findings in the study of lattice results of the $DK$ system by means of an auxiliary potential of the type of Eq. (\ref{eq:vcdd}), which demanded $\sqrt{s_{\mathrm{CDD}}}$ to be more than $300$ MeV above the $DK$ threshold \cite{sasa}. The results that we obtain, using values of $\sqrt{s_{\mathrm{CDD}}}$ bigger than $M_K+M_{K^*}+300$ MeV, barely change the results that we have shown. Smaller values of the mass excess bring changes in the upper part of $M_\mathrm{inv}$ in the plot but not very close to the threshold.
By comparing Figs. \ref{fig:fig5} and \ref{fig:fig7} we observe the following features.
For $Z=0$, both methods are identical. For $Z\neq 0$ the results with both methods change a bit, with the differences bigger than with the change of the cutoff. The differences become bigger as $Z$ increases. Yet, for $Z=0.2$ and close to threshold the results vary from $9$ to $8.3$, an $8$ \% change, or $\pm 4$ \% from the average of the two methods. For $Z=0.6$ the differences are bigger, from $6.5$ to $5$ or $\pm 13$ \% from the average. From this study and the value
of $dR_\Gamma/dM_{\mathrm{Inv}}$ at threshold we find that the uncertainties of $Z$ are of the order of $0.1$. 

So far, we have only discussed the ratio close to threshold. However, the invariant mass dependence is also useful to determine $Z$. The method of the CDD pole is more general since it has one extra parameter, $s_{\mathrm{CDD}}$. In particular for larger values of $s_{\mathrm{CDD}}$ one would encounter the potential of Eq. (\ref{eq:lpot}) and conclude that a CDD pole is unnecesary. One could make a fit to the spectrum and determine $Z$ and $s_{\mathrm{CDD}}$, from where one gets extra information about the nature of the resonance. 

It is interesting to quote here what we get for $Z$ using the chiral unitary approach of Section \ref{sec:chroca}. By means of the coupling obtained there and using the cutoff of $950$ MeV, demanded to fix the pole at $\sqrt{s_R}=1285$~MeV, we obtain $Z=0.57$, very similar to the value $Z=0.5$ in Ref.~\cite{geng}. In this case the distribution of $dR_\Gamma/dM_{\mathrm{Inv}}$ is shown in Fig. \ref{fig:fig4} and is undistinguible from what we obtain by using the linear potential and Eq. (\ref{eq:rat3}) with this value of $Z$. As to the meaning of this value, stemming from the energy dependence of the potential of Eq. (\ref{eq:rocap}), it is unclear whether it reflects missing meson-meson channels or some possible elementary component. What matters for the present work is the accuracy of the method presented here to determine $Z$ from the $dR_\Gamma/dM_{\mathrm{Inv}}$ magnitude. The idea to divide $d\Gamma$ by the phase space to see the shape of the resonance has also being exploited experimentally in Ref.
\cite{delAmoSanchez:2010yp} in the $B^+\to K^+ K^- \pi^+$ reaction to see the tail of the $f_0(980)$ resonance from the $K^+K^-$ spectrum. See
also the related theoretical paper of Ref.~\cite{Dias:2016gou}.
\section{Direct production of the elementary components}
\begin{figure}[t]
 \begin{center}
  \includegraphics[width=1\linewidth]{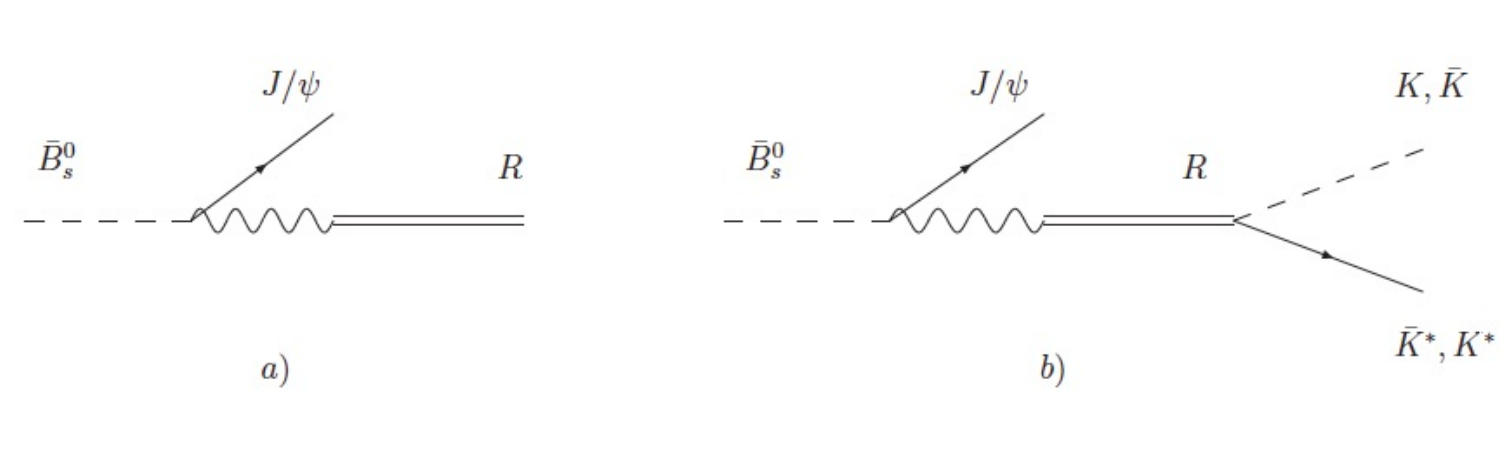}
 \end{center}
\caption{a) Direct production of the ``elementary'' component of the resonance. b) Contribution to $K\bar{K}^*$ production form the ``elementary'' components.}
\label{fig:fig9}
\end{figure}

So far, we have assumed that the resonance formation and the $K\bar{K}^*$ formation proceed via the mechanism of Figs. \ref{fig:fig2} and \ref{fig:fig3}. Yet, if we have a non-$K\bar K^*$ component we can also think of producing  it directly in the weak decay. This means that we must include the new mechanisms depicted in Fig. \ref{fig:fig9}. This follows the line of argumentation of Ref. \cite{pedrozhifeng}.
In Fig. \ref{fig:fig9} we have two unknowns, the coupling of $\bar{B}^0_s$ to $J/\psi$ plus the ``elementary'' component, and the coupling of the resonance to this component. We choose to parametrize this contribution as a fraction, $\delta$, of the former one, such that the new mechanism for the coalescence will give a contribution
\begin{equation}
 t'(\bar{B}^0_s\to J/\psi R)= V_p G(s_R) g\,\delta\ .
\end{equation}
As a consequence, the contribution to the $K\bar{K}^*$ production process of Fig. \ref{fig:fig9}b) will be
\begin{eqnarray}
&& t'(\bar{B}^0_s\to J/\psi K\bar{K}^*)=V_P G(s_R)g\,\delta\frac{1}{s-s_R}g\nonumber\\&&=V_PG(s_R)\,\delta \,t_{K\bar{K}^*}\nonumber\\
\end{eqnarray}
where in the last step we have used the pole approximation to the amplitude, $t_{K\bar{K}^*,K\bar{K}^*}\simeq g^2/(s-s_R)$.
Hence, the numerator in Eqs. (\ref{eq:rat3}), (\ref{eq:rat4}) is changed to
\begin{eqnarray}
 1+Gt\to 1+Gt+G(s_R)\,\delta t=1+(G(s)+G(s_R)\delta) t\ ,\nonumber\\
\end{eqnarray}
while the denominator is changed as 
\begin{equation}
 gG(s_R)\to gG(s_R)(1+\delta)\ .
\end{equation}
Thus the factor of Eq. (\ref{eq:rat3}) or Eq. (\ref{eq:rat4}) becomes now
\begin{eqnarray}
 &&\left|\frac{1+Gt}{gG_R}\right|^2\to-\left.\frac{\partial G}{\partial s}\right|_{s_R}\frac{1}{1-Z}\left|\frac{1}{1+\delta}\right|^2\times\nonumber\\&&\left|G(s_R)^{-1}+(G(s_R)^{-1}G(s)+\delta)t\right|^2\nonumber\\
\label{eq:rat5}
\end{eqnarray}
with
\begin{equation}
 t=\frac{1}{V^{-1}-G(s)}
\end{equation}
and
\begin{eqnarray}
 &&V=\nonumber\\&&G(s_R)^{-1}+\left.\frac{\partial G}{\partial s}\right|_{s_R}G(s_R)^{-2}\frac{Z}{1-Z}\frac{(s_R-s_{\mathrm{CDD}})(s-s_R)}{(s-s_{\mathrm{CDD}})}\ .\nonumber\\\label{eq:potdel}
\end{eqnarray}
It might look that Eq. (\ref{eq:rat5}) does not have the nice property of Eq. (\ref{eq:rat4}) with respect to the cutoff invariance. Actually, following the same
steps that led to this latter equation one can readily see that the new expression in Eq. (\ref{eq:rat5}) can be recast like Eq. (\ref{eq:rat4}) simply replacing the $1$ in the 
numerator of the last factor by
\begin{eqnarray}
 &&1+VG(s_R)\delta=\nonumber\\&&\left.1+\delta\left(1+\frac{\partial G}{\partial s}\right|_{s_R}G(s_R)^{-1}\frac{Z}{1-Z}\frac{(s_R-s_{\mathrm{CDD}})(s-s_R)}{s-s_{\mathrm{CDD}}}\right)\ .\nonumber\\\label{eq:del1}
\end{eqnarray}
Now $\left.\frac{\partial G}{\partial s}\right|_{s_R}G(s_R)^{-1}$ depends on the cutoff, but we can reunify
\begin{equation}
 \left.\frac{\partial G}{\partial s}\right|_{s_R}G(s_R)^{-1}=\left.\frac{\partial }{\partial s}\mathrm{ln} G(s)\right|_{s_R}
\end{equation}
and then $\mathrm{ln}G(s)$ is a soft function of the cutoff. Furthermore, the second term in the bracket multiplying $\delta$ in Eq. (\ref{eq:del1}) is reasonably smaller than the unity preceding it. All this guarantees a smooth cutoff dependence of the new term. But more important, changes induced by changes in the cutoff in Eq. (\ref{eq:rat5}) close to threshold can be incorporated by changing $\delta$, and since $\delta$ is unknown and will be changed to see the stability of the results, by performing this test we implicitly accommodate the cutoff dependence of the results.

\begin{figure*}[htb]
 \begin{center}
  \includegraphics[width=1\linewidth]{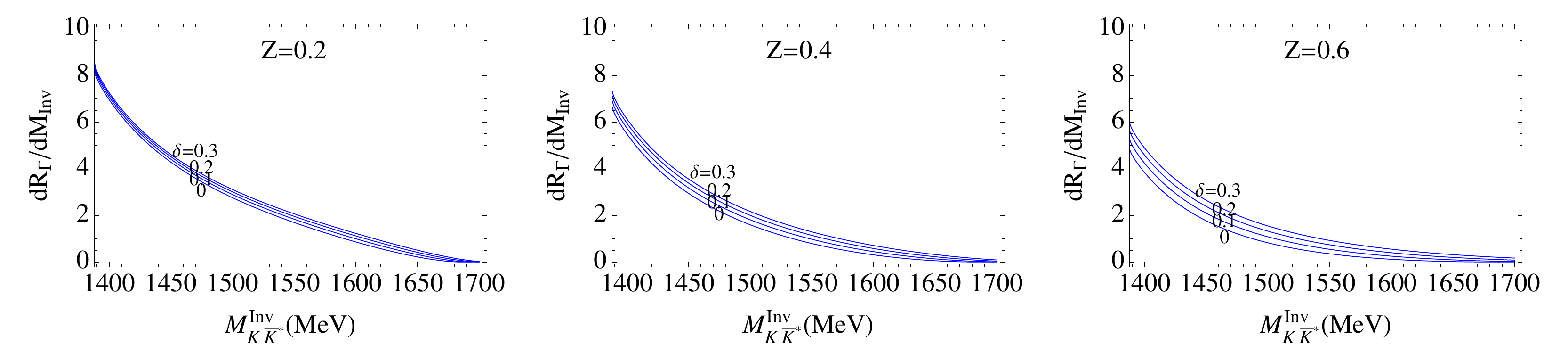}
 \end{center}
\caption{$d R_\Gamma/d M_\mathrm{inv}$ from Eq. (\ref{eq:rat5}) as a function of $M_\mathrm{inv}$ for several $Z$ and $\delta$ values and $q_\mathrm{max}=950$ MeV.}
\label{fig:delz}
\end{figure*}

In Fig. \ref{fig:delz} we show the results obtained by using $\delta=0,0.1,0.2$ and $0.3$ fractions. Note that $(1+\delta)^2$ for $\delta=0.3$ already introduces a $70$ \% increase in the rate for the coalescence process. Since the idea is to apply the present method for cases where we have large molecular components, such values of $\delta$ are reasonable. Then, in order to quantify uncertainties from this new mechanism, we evaluate again $dR_\Gamma/dM_{\mathrm{inv}}$ including the new 
corrections. We do the exercise using the CDD version of the potential. 
\begin{figure}[htb]
 \begin{center}
  \includegraphics[width=1\linewidth]{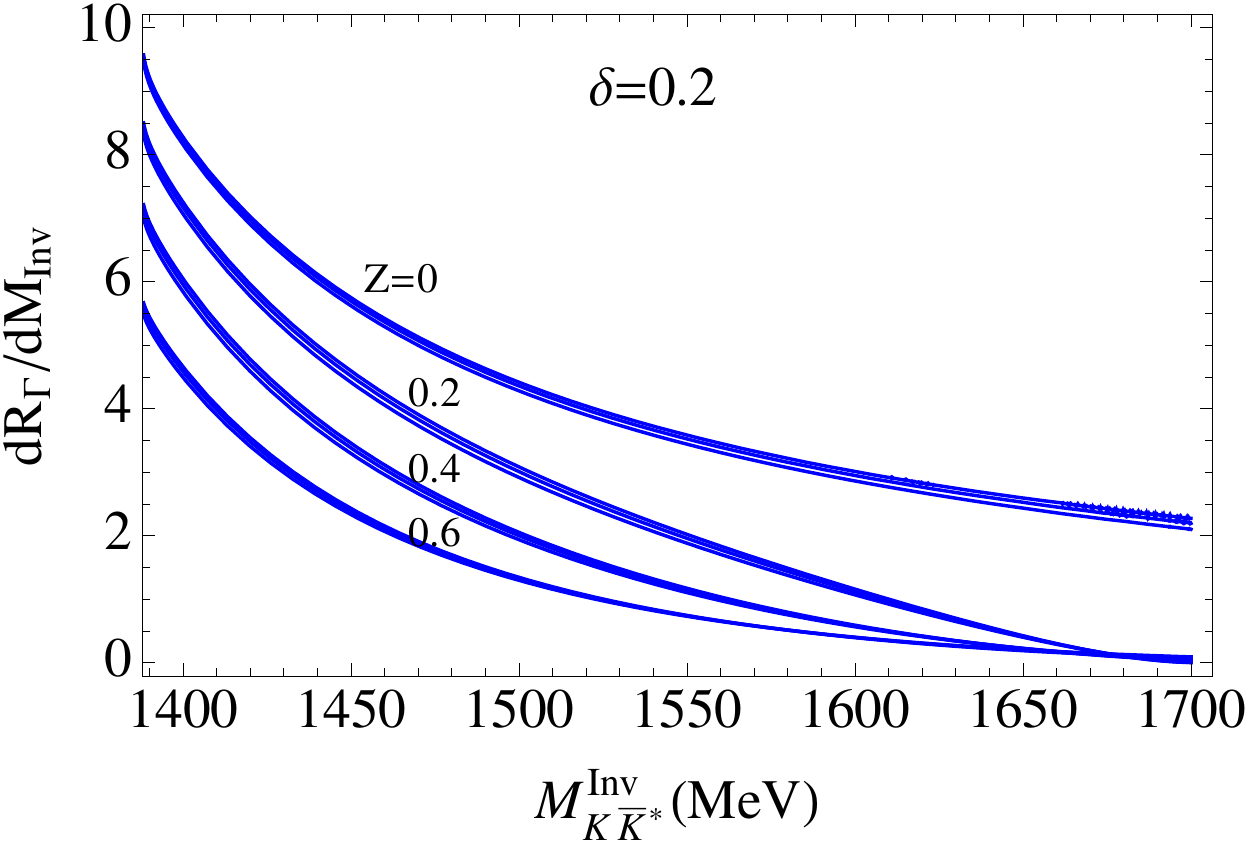}
 \end{center}
\caption{$d R_\Gamma/d M_\mathrm{inv}$ from Eq. (\ref{eq:rat5}) as a function of $M_\mathrm{inv}$ for several $Z$ and different values of the cutoff, $q_\mathrm{max}=850,950$ and $1050$ MeV. $\delta$ is fixed to $0.2$.}
\label{fig:delz2}
\end{figure}

In Fig. \ref{fig:delz} we see the results for $dR_\Gamma/dM_{\mathrm{Inv}}$ for $Z=0.2,0.4$ and $Z=0.6$ and different values of $\delta$ ($q_\mathrm{max}=950$ is used)\footnote{Note that for $Z=0$ the result coincides with Fig. \ref{fig:fig7}.}. We see small changes by varying $\delta$. For $Z=0.2$ the changes are moderate. They are a bit bigger for $Z=0.4$ and even bigger for $Z=0.6$. But even for $Z=0.6$ the results at threshold from $\delta=0$ to $\delta=0.2$ vary only by about $10$\%. The reason for this stability is that the ratio with the new mechanisms alone is also of the type of $g^2/(s-s_R)^2$ as in Eq. (\ref{eq:coup}), so the mechanisms are practically 
proportional and the ratio is maintained. To observe the uncertainties in the cutoff, in Fig. \ref{fig:delz2} we show the ratio from Eq. (\ref{eq:rat5}) for several $Z$ and different values of the cutoff, $q_\mathrm{max}=850,950$ and $1050$ MeV. $\delta$ is fixed to $0.2$. We can see that the curves barely move as commented above.

Altogether, and summing different sources of uncertainties in quadratures, the conclusions that we drew before hold also now and one can obtain with this method the value of $Z$ with uncertainties of about $\pm 0.1$.

  Let us note that through the derivations done we have always assumed that we have the $C=+$ state of $K \bar K^*$ in the final state. Actually, the $C=-$ combination is also possible. The test conducted demands that we isolate the positive $C$-parity in the final state. This is in principle possible experimentally as shown in Refs. \cite{aston,abele}. Certainly it would help if the process is dominated by the $f_1(1285)$. Nevertheless, the study carried here is quite general and can be applied to multiple cases where one suspects that one resonance has large components of  a composite system.

\section{Conclusions}
We propose a new method to determine the amount of compositeness of resonances mostly made from two hadron components, and bound with respect to these components. The method relies on the comparison of two independent but related quantities. On the one hand, one measures the production rate of the resonance in some reaction, independently of the decay channel of the resonance.  On the other hand, the mass distribution of the two components close to threshold is measured, which appears above the resonance energy since the resonance is supposed to be a bound state of these components. The resonance tail appears as an enhancement in the invariant mass distribution close to threshold, easily distinguishable from a pure phase space distribution. The method consists of taking the ratio of this mass distribution over the rate for the resonance production in the same reaction, also removing the phase space factors. We show that the strength is related to the coupling of the resonance to the two hadron component, and the shape to the square of the resonant amplitude. The ratio decreases as a function of the invariant mass at threshold. Then, a method is used to relate the measured ratio to the compositeness of the resonance. We show that it is possible to determine the compositeness, 1-Z, from that ratio. Systematic uncertainties are estimated to conclude that it is possible to measure values of Z with about 0.1 of uncertainty. We used a particular case for the numerical evaluations in the decay of $B^0_s \to J/\psi f_1(1285)$ which has a large composite component of $K \bar K^*$. Yet, the method is general and can be applied to any case where one has hints, or theoretical backing, for the molecular nature of a resonance. In view of this, we encourage the simultaneous measurement of these quantities, that for the moment appear in different reactions, or are measured by different groups, and usually are given in terms of counts and not absolute measurements, as is demanded by the method proposed.

\section{Acknowledgements}

This research has been supported by
the Spanish Ministerio de Econom\'ia y Competitividad (MINECO), European FEDER funds under the contracts FIS2014-51948-C2-2-P and SEV-2014-0398, by Generalitat Valenciana under Contract PROMETEOII/2014/0068. Also, this work is supported by the National Science Foundation (CAREER grant No. 1452055, PIF grant No. 1415459), and by
GWU (startup grant).


\begin{thebibliography}{99}
\bibitem{Klempt:2007cp} 
  E.~Klempt and A.~Zaitsev,
  Phys.\ Rept.\  {\bf 454}, 1 (2007)

\bibitem{Crede:2008vw} 
  V.~Crede and C.~A.~Meyer,
  Prog.\ Part.\ Nucl.\ Phys.\  {\bf 63}, 74 (2009)

\bibitem{Klempt:2009pi} 
  E.~Klempt and J.~M.~Richard,
  Rev.\ Mod.\ Phys.\  {\bf 82}, 1095 (2010)

\bibitem{npa} 
  J.~A.~Oller, E.~Oset and J.~R.~Pelaez,
  Phys.\ Rev.\ D {\bf 59}, 074001 (1999)
  [Phys.\ Rev.\ D {\bf 60}, 099906 (1999)]
  [Phys.\ Rev.\ D {\bf 75}, 099903 (2007)]

\bibitem{kaiser} 
  N.~Kaiser,
  Eur.\ Phys.\ J.\ A {\bf 3}, 307 (1998).

\bibitem{locher} 
  M.~P.~Locher, V.~E.~Markushin and H.~Q.~Zheng,
  Eur.\ Phys.\ J.\ C {\bf 4}, 317 (1998)

\bibitem{juanarriola} 
  J.~Nieves and E.~Ruiz Arriola,
  Nucl.\ Phys.\ A {\bf 679}, 57 (2000)

\bibitem{cola} 
  D.~Jido, J.~A.~Oller, E.~Oset, A.~Ramos and U.~G.~Meissner,
  Nucl.\ Phys.\ A {\bf 725}, 181 (2003)

\bibitem{Borasoy:2005ie} 
  B.~Borasoy, R.~Nissler and W.~Weise,
  Eur.\ Phys.\ J.\ A {\bf 25}, 79 (2005)

\bibitem{Oller:2005ig} 
  J.~A.~Oller, J.~Prades and M.~Verbeni,
  Phys.\ Rev.\ Lett.\  {\bf 95}, 172502 (2005)

\bibitem{Oller:2006jw} 
  J.~A.~Oller,
  Eur.\ Phys.\ J.\ A {\bf 28}, 63 (2006)

\bibitem{Borasoy:2006sr} 
  B.~Borasoy, U.-G.~Meissner and R.~Nissler,
  Phys.\ Rev.\ C {\bf 74}, 055201 (2006)

\bibitem{Hyodo:2008xr} 
  T.~Hyodo, D.~Jido and A.~Hosaka,
  Phys.\ Rev.\ C {\bf 78}, 025203 (2008)

\bibitem{Mai:2014xna} 
  M.~Mai and U.~G.~Meißner,
  Eur.\ Phys.\ J.\ A {\bf 51}, no. 3, 30 (2015)


\bibitem{reviewreso} 
  J.~A.~Oller, E.~Oset and A.~Ramos,
  Prog.\ Part.\ Nucl.\ Phys.\  {\bf 45}, 157 (2000)

\bibitem{compositeness} 
  S.~Weinberg,
  Phys.\ Rev.\  {\bf 137}, B672 (1965).

\bibitem{kalash} 
  V.~Baru, J.~Haidenbauer, C.~Hanhart, Y.~Kalashnikova and A.~E.~Kudryavtsev,
  Phys.\ Lett.\ B {\bf 586}, 53 (2004)


\bibitem{danijuan} 
  D.~Gamermann, J.~Nieves, E.~Oset and E.~Ruiz Arriola,
  Phys.\ Rev.\ D {\bf 81}, 014029 (2010)

\bibitem{rocasingh} 
  L.~Roca, E.~Oset and J.~Singh,
  Phys.\ Rev.\ D {\bf 72}, 014002 (2005)
\bibitem{Zhou:2014ila} 
  Y.~Zhou, X.~L.~Ren, H.~X.~Chen and L.~S.~Geng,
  Phys.\ Rev.\ D {\bf 90}, no. 1, 014020 (2014)



\bibitem{Aaij:2013rja} 
  R.~Aaij {\it et al.} [LHCb Collaboration],
  Phys.\ Rev.\ Lett.\  {\bf 112}, no. 9, 091802 (2014)

\bibitem{Liang:2014tia}
W.~H.~Liang and E.~Oset,
Phys.\ Lett.\ B {\bf 737}, 70 (2014)

\bibitem{Bramon:1992kr} 
  A.~Bramon, A.~Grau and G.~Pancheri,
  Phys.\ Lett.\ B {\bf 283}, 416 (1992).

\bibitem{hanhart} 
  J.~T.~Daub, C.~Hanhart and B.~Kubis,
  JHEP {\bf 1602}, 009 (2016)
  
\bibitem{weiwang} 
  W.~F.~Wang, H.~n.~Li, W.~Wang and C.~D.~Lü,
  Phys.\ Rev.\ D {\bf 91}, no. 9, 094024 (2015)


\bibitem{Kang:2013jaa} 
  X.~W.~Kang, B.~Kubis, C.~Hanhart and U.~G.~Meißner,
  Phys.\ Rev.\ D {\bf 89}, 053015 (2014)

\bibitem{Liang:2015twa} 
  W.~H.~Liang, J.~J.~Xie, E.~Oset, R.~Molina and M.~Döring,
  Eur.\ Phys.\ J.\ A {\bf 51}, no. 5, 58 (2015)


\bibitem{birse} 
  M.~C.~Birse,
  Z.\ Phys.\ A {\bf 355}, 231 (1996)

\bibitem{hidden1} 
  M.~Bando, T.~Kugo, S.~Uehara, K.~Yamawaki and T.~Yanagida,
  Phys.\ Rev.\ Lett.\  {\bf 54}, 1215 (1985).

\bibitem{hidden2} 
  M.~Bando, T.~Kugo and K.~Yamawaki,
  Phys.\ Rept.\  {\bf 164}, 217 (1988).

\bibitem{hidden4} 
  U.~G.~Meissner,
  Phys.\ Rept.\  {\bf 161}, 213 (1988).





\bibitem{yamagata} 
  J.~Yamagata-Sekihara, J.~Nieves and E.~Oset,
  Phys.\ Rev.\ D {\bf 83}, 014003 (2011)

\bibitem{aceti} 
  F.~Aceti and E.~Oset,
  Phys.\ Rev.\ D {\bf 86}, 014012 (2012)
\bibitem{jido} 
  T.~Hyodo, D.~Jido and A.~Hosaka,
  Phys.\ Rev.\ C {\bf 85}, 015201 (2012)

\bibitem{hyodo} 
  T.~Hyodo,
  Int.\ J.\ Mod.\ Phys.\ A {\bf 28}, 1330045 (2013)


\bibitem{sekihara} 
  T.~Sekihara, T.~Hyodo and D.~Jido,
  PTEP {\bf 2015}, 063D04 (2015)


\bibitem{juancarmen} 
  C.~Garcia-Recio, C.~Hidalgo-Duque, J.~Nieves, L.~L.~Salcedo and L.~Tolos,
  Phys.\ Rev.\ D {\bf 92}, no. 3, 034011 (2015)

\bibitem{acetidai} 
  F.~Aceti, L.~R.~Dai, L.~S.~Geng, E.~Oset and Y.~Zhang,
  Eur.\ Phys.\ J.\ A {\bf 50}, 57 (2014)

\bibitem{castillejo} 
  L.~Castillejo, R.~H.~Dalitz and F.~J.~Dyson,
  Phys.\ Rev.\  {\bf 101}, 453 (1956).

\bibitem{nsd} 
  J.~A.~Oller and E.~Oset,
  Phys.\ Rev.\ D {\bf 60}, 074023 (1999)

\bibitem{sasa} 
  A.~Mart\'inez Torres, E.~Oset, S.~Prelovsek and A.~Ramos,
  JHEP {\bf 1505}, 153 (2015)
\bibitem{geng}
  L.~S.~Geng, X.~L.~Ren, Y.~Zhou, H.~X.~Chen and E.~Oset,
Phys.\ Rev.\ D {\bf 92}, no. 1, 014029 (2015)
\bibitem{delAmoSanchez:2010yp} 
  P.~del Amo Sanchez {\it et al.} [BaBar Collaboration],
  Phys.\ Rev.\ D {\bf 83}, 052001 (2011)
\bibitem{Dias:2016gou}
J.~M.~Dias, F.~S.~Navarra, M.~Nielsen and E.~Oset,
arXiv:1601.04635 [hep-ph].
\bibitem{pedrozhifeng} 
  Z.~F.~Sun, M.~Bayar, P.~Fernandez-Soler and E.~Oset,
  Phys.\ Rev.\ D {\bf 93}, no. 5, 054028 (2016)


\bibitem{aston} 
  D.~Aston {\it et al.},
  Phys.\ Lett.\ B {\bf 201}, 573 (1988).


\bibitem{abele} 
  A.~Abele {\it et al.} [Crystal Barrel Collaboration],
  Phys.\ Lett.\ B {\bf 415}, 280 (1997).


\end{thebibliography}
\end{document}